\documentstyle[12pt]{article}
\begin{document}
\hskip 9cm HUPD9317
\par
\hskip 9cm YAMAGATA-HEP-93-12
\par
\hskip 9cm IPS Res. Rep. No.93-09
\par
\vskip 2cm
\centerline{\LARGE Scaling Study of Pure Gauge Lattice QCD by}
\par
\vskip 5mm
\centerline{\LARGE Monte Carlo Renormalization Group Method}
\par
\vskip 15mm
\centerline{\large QCDTARO Collaboration}
\par
\vskip 1cm
\centerline{\bf
K.Akemi, M.Fujisaki, M.Okuda, Y.Tago
}\par
\centerline{
Research Center for Computational Science,
Fujitsu Limited,
}\par
\centerline{
Mihama-ku, Chiba 261, Japan
}\par
\vskip 1cm
\centerline{\bf
Ph de Forcrand
}\par
\centerline{
IPS, ETH-Z\"urich, CH-8092 Z\"urich, Switzerland
}\par
\vskip 1cm
\centerline{\bf
T.Hashimoto
}\par
\centerline{
Department of Applied Physics, Faculty of Engineering,
}\par
\centerline{
Fukui University, Fukui 910, Japan
}\par
\vskip 1cm
\centerline{\bf
S.Hioki, O.Miyamura, T.Takaishi
}\par
\centerline{
Department of Physics, Hiroshima University,
}\par
\centerline{
Higashi-Hiroshima 724, Japan
}\par
\vskip 1cm
\centerline{\bf
A.Nakamura
}\par
\centerline{
Faculty of Education, Yamagata University,
}\par
\centerline{
Yamagata 990, Japan
}\par
\vskip 1cm
\centerline{\bf
I.O.Stamatescu
}\par
\centerline{
FEST Heidelberg and Institut f\"ur Theoretische Physik
}\par
\centerline{
Universit\"at  Heidelberg, D-6900 Heidelberg, Germany
}
\par
\vskip 1cm
{\bf Abstract}
\par
The scaling behavior of pure gauge SU(3) in the region $\beta=5.85 - 7.60$
is examined by a Monte Carlo Renormalization Group analysis.
The coupling shifts induced by factor 2
blocking are measured both on 32$^4$ and 16$^4$ lattices
with high statistics.
A systematic deviation from naive 2-loop scaling is
clearly seen.
The mean field and effective coupling constant schemes
explain part, but not all of the deviation.
It can be accounted for by a suitable
change of coupling constant, including a correction term
${\cal O}(g^7)$ in the 2-loop lattice $\beta$-function.
 Based on this improvement,
$\sqrt{\sigma}/\Lambda_{\overline {MS}}^{n_f=0}$ is
estimated to be $2.2(\pm 0.1)$ from the analysis of the
string tension $\sigma$.
\par
\pagebreak
Since a confirmation of the approach to the continuum limit has
basic importance
in lattice Quantum Chromo Dynamics (lattice QCD),
systematic scaling analyses at larger $\beta$ are
inevitably required. Recent analyses have shown that scaling violations
persist in physical quantities such as the string tension
and hadron masses up to $\beta=6.8$.\cite{ref1}
On this problem, it has been argued that
a suitably chosen coupling constant reveals perturbative scaling.
\cite{ref2,ref3,ref4,ref5,ref6} \par
Monte Carlo Renormalization Group (MCRG) tells
us the coupling shift $\Delta\beta$ induced by scale transformation
of the lattice spacing, $a \rightarrow sa$;$^7$\par
$$
   s= exp[\int^{g(\beta-\Delta\beta)}_{g(\beta)} {dg\over \beta_f(g)}]
    = {f({1\over g(\beta-\Delta\beta)^2})\over f({1\over g(\beta)^2})} \eqno(1)
$$
where $f(1/g^2)=a\Lambda$ and $\beta_f(g)$ is the lattice $\beta$ function.
(For bare lattice coupling constant, $1/g^2=\beta/6$.)
This gives us another way to examine  the scaling behavior of lattice QCD.
In SU(3) lattice gauge systems,
MCRG has been performed by several groups on 16$^4$ lattices
in the large $\beta$ region up to 7.2.\cite{ref8,ref9,ref10,ref11,ref12}
However, these results were not conclusive and even controversial.
Although Gupta et al.\cite{ref11} have claimed consistency with
asymptotic scaling from their $\sqrt 3$ blocking result,
Bowler et al.\cite{ref9,ref10} have found
sizeable deviations from 2-loop scaling.  Recently Hoek has reanalyzed
the same data and claimed a very slow approach to  scaling.\cite{ref12}
These varied conclusions are mainly
due to the difficulty of obtaining a precise value for the coupling shift
in high $\beta$ region, because the deconfining transition prevents an
accurate matching between Wilson loops on a small lattice.
Studies on larger
lattices with better statistics are required to clarify the scaling
behavior in the high $\beta$ region. \par
In this work, we report results of an MCRG study and scaling analysis
in the high $\beta$ region
both on 32$^4$ and 16$^4$ lattices with high statistics.  \par
Using a 32$^4$ lattice gives us the following advantages.\par
i) one more blocking level than previous works. It is not apriori clear that
blocking from 16$^4$ to 2$^4$ is
sufficient for loop matching.
A deeper blocking is preferred to confirm good matching
and for closer matching conditions.
\par
ii) we remain in the confinement phase up to
$\beta \sim 6.9$ whereas a 16$^4$ lattice is above the deconfining transition
 point for $\beta>6.35$.\par
\vskip 10mm

The numerical simulation has been performed on 512-cell parallel processor
Fujitsu AP1000.\cite{ref13}
Three lattices of size 32$^4$, 16$^4$ and 8$^4$ are generated by
the over-relaxed pseudo heatbath algorithm\cite{ref14}
 (the mixing ratio of over-relaxation to
pseudo heat bath is 9:1 in average).
Blocking is performed every 10 updates.
2K - 3K configurations are blocked at each $\beta$ (more near the deconfining
transition point).
For error estimation, the jack-knife method is applied.
We monitor autocorrelations of blocked Wilson loops in those measurements
to know the statistical validity of sampling.(See Fig.3)
Near the deconfining transition point, twice more configurations are
blocked.  Effect of long autocorrelation there is taken into account in the
error
estimation. Details of prescription will be presented elsewhere.

In this work, the Swendsen blocking transformation is used to
double the lattice spacing($s=2$)\cite{ref15}.
Blocking is repeatedly performed down to $2^4$ for two lattices,
one of size $L$ ( at $\beta$) and
the other of size $L/2$ (at $\beta -\Delta \beta$).
To match long range physical contents on both lattices,
a set of Wilson loops on one blocked lattice,
is compared with the corresponding one on the other blocked lattice.
For early matching between blocking trajectories,
the blocking transformation is controled by a parameter $q$ which governs
the size of Gaussian fluctuations around the maximal SU(3) projection of
the block link variable\cite{ref16}.
The coupling shift $\Delta \beta$ is determined at the value of $q$ where
the mismatch of the two sets of Wilson loops is minimum.
Planar $1\times 1$ and $1\times 2$ Wilson loops,
and non-planar 6-link ("twist" and "chair") and 8-link ("sofa") loops
are measured on blocked lattices at $q= 0.0, 0.02, 0.04$ for this purpose
(see Fig.1).
A brief description is found in our previous report\cite{ref13}.
\par
\vskip 10mm

$\Delta\beta$ at $\beta=5.85-7.60$ measured by matching 16$^4$ and
8$^4$ lattices after three
blocking steps is shown by open circles
in Fig.2a.  Improved statistics give a clear systematic behavior for the new
data in comparison with those of previous works (Fig.2b)
although they are consistent.
The data are significantly below the 2-loop  scaling result(solid curve).
Therefore naive 2-loop  scaling does not hold in this region.
The deviation rapidly decreases as $\beta$ increases. But,
even at $\beta=7.60$, $10\%$ deviation remains.\par
A notable feature of the present data is the following.
For $\beta < 6.3$ , the quality of the data
is sufficient and shows the approach to the
2-loop  scaling result. On the other hand,
the data above $\beta =6.3$ suffer relatively large errors.
The matching of blocked Wilson
loops becomes difficult in this region. This difficulty
is caused by the deconfining transition on the 16$^4$ lattice.\cite{ref17}
As shown in
Fig.3, the autocorrelation time of blocked Wilson loops
sharply peaks at $\beta = 6.35\pm 0.05$ ( On $8^4$ lattice, $\beta=5.90$).
An increase of fluctuations
of the blocked Wilson loops is also evident at this point.
Thus, the 16$^4$ lattice turns into deconfining region
at this point.
It is noted that the value $\beta=6.35$ is smaller than the previously
expected value $6.45 \pm 0.05$ from an analysis of the
finite temperature phase transition\cite{ref18}. Above this $\beta$,
Wilson loops are dominated by perturbative contributions and the matching
suffers large errors.
\par
\vskip 10mm

Matching between 32$^4$ and 16$^4$ has been tried at $\beta=6.35 - 7.00$.
The deconfining transition point
on a 32$^4$ lattice is pushed up to around $\beta=6.9$ since the coupling
shift is $\approx 0.54$ in this region and the deconfining transition
point is $\beta=6.35$ on a 16$^4$ lattice.
At $\beta$=6.35,6.55,6.65 and 6.80 where the lattice is in the confinement
phase,
 the matching can be performed successfully
and the resultant coupling shifts after 3 and 4 blockings agree within
error as shown in Fig.4. Thus,the data become stable for blockings greater than
three and this fact gives reliability to the 16$^4$ lattice
(three blockings) measurements.\par
The measured coupling shifts (black circles in Fig.2a)
are naturally connected with those of $\beta<6.35$ on the 16$^4$ lattice.
Thus, a systematic deviation from naive 2-loop  scaling
in this region is further confirmed by these measurements.
\par
At $\beta=7.00$ where the 32$^4$ lattice is above the deconfining transition
point, definite data could not be extracted from our 3K blocked configurations
(30K sweeps) due to
very long range fluctuations.
\par
\vskip 10mm

Although a trend to approach the 2-loop scaling value is seen,
the coupling shift
shows significant deviation. This deviation can be partly
absorbed by a mean field scheme\cite{ref2,ref3} or
an effective coupling constant scheme.\cite{ref4,ref5,ref6}
In the mean field scheme, the $\overline {MS}$ coupling constant is given by
 $1/g_{\overline {MS}}(\pi/a)^2= <U_{plaq}>/g^2 + 0.025$  where $<U_{plaq}>$ is
the average plaquette.  Similarly the effective coupling constant is defined as
$g_e^2=3(1-<U_{plaq}>)$.
Both coupling constants are obtained here based on our measurement of
$<U_{plaq}>$ at $\beta=5.70-7.60$.
The coupling shift is calculated assuming the 2-loop scaling
form for $g_{\overline{MS}}$\cite{ref19}
$$
   f_2(x)=\pi({x+({b_1\over b_0})\over b_0})^{b_1\over 2b_0^2}
exp(-{x\over 2b_0}),\ \ \ \
x={1\over g_{\overline{MS}}^2}
  \eqno(2)
$$
where $b_0=33/(48\pi^2)$  and $b_1=(102/121)b_0^2$.
As shown by  curve M in Fig.5,
this scheme partly explains the present MCRG data.
Similary, the effective coupling constant $g_e$ improves
the agreement somewhat, as shown by curve E in Fig.5.
\par
However, a sizeable deviation still persists between these schemes and the
present data. Therefore we need a different prescription to approach
the continuum limit from the presently accessible region of $\beta$.
In ref.2,3, to get the continuum limit, a linear or bilinear extrapolation
in terms of the lattice constant $a$ was assumed for
physical quantities expressed in unit of $\Lambda_{\overline{MS}}$,
 while the authors of ref.6
used a linear extrapolation in $1/lna$.
Here instead, the continuum limit is extracted by taking into account the
lattice $\beta$ function and expressing our MCRG data in terms of an effective
coupling constant.
We fit the data by
the lattice $\beta$-function including a next-order correction, and define
our effective coupling $1/g_u^2$ by a free shift of
$1/g_{\overline{MS}}^2$
as follows:
$$
  - {dx_u\over 2dlna}= b_0 + {b_1\over x_u} + {b'\over x_u^2},     \eqno(3a)
$$
with
$$
 x_u \equiv {1\over g_u^2} ={1\over g_{\overline {MS}}^2}-x_0\ \ \ \ . \eqno(4)
$$
Assuming the correction term $b'$ is small,
we actually use the following equation instead of eq.(3a),
$$
- {dx_u\over 2b_0}[{1\over (1 + {b_1\over b_0x_u})} - {b'\over b_0x_u^2}]=dlna.
    \eqno(3b)
$$
The solution of eq.(3b) is

$$
         f(x_u)=f_2(x_u)exp({-b'\over 2b_0^2x_u}) \ \ \ .  \eqno(5)
$$
Using eqs.(4) and (5),
we have attempted two fits for the data above  $\beta = 6.00$ :
(A) Restrict $x_u=1/g_{\overline{MS}}^2$,
(i.e.$x_0=0$) with $b'$ as a free parameter,
(B) Allow a shift of the effective coupling constant,
i.e. both $x_0$ and $b'$ are free parameters.
The results are shown in Fig.5. In case (A),
only a poor fit is obtained.
The coefficient $b'$ is also relatively large as
$b'/b_0=-0.0850(0.0013)$ while $b_1/b_0=0.0587$.
Thus, it is difficult to explain the present data by
an additional $O(g_{\overline {MS}}^7)$ term only.
On the other hand in case (B), we can fit the data quite well by
values of parameters as (curve B)
$$
   x_0=0.442(0.004),\ \ \ \ \ {b'\over b_0}=-0.0119(0.0008) \ \ \ .\eqno(6)
$$
In this case, the coefficient of the $O(g_u^7)$ term is small but
the shift parameter $x_0$ is non-negligible. It is noted that
we have tried fitting in the effective coupling scheme also and get
similar results. Since the sign of $b'$ is negative, we say that
our lattice is coarser than predicted by 2-loop scaling.
\par
Based on this scheme, we can discuss the continuum limit of physical
quantities in units of $\Lambda_{\overline{MS}}^{n_f=0}$.  In Fig.6, scaling of
the string tension of ref.s 6,20,21 is examined.  As expected,
good scaling is obtained for $\sqrt{\sigma}/\Lambda_{\overline{MS}}^{n_f=0}$
for $\beta$=5.7-6.8 as shown by black circles in the figure and the value is
$$
\sqrt{\sigma }/\Lambda_{\overline{MS}}^{n_f=0}=2.2(0.1). \eqno(7)
$$
This value is also interpreted as the continuum limit value in the present
scheme.
Our result (7)
is larger than that of
the static quark potential in ref.6 ($1.80\pm 0.06$) and ref.22 ($1.72\pm
0.13$).
We note also that
$\Lambda_{\overline{MS}}^{n_f=0}=200MeV$ for $\sqrt{\sigma}=440 MeV$.
It
is slightly smaller than that
extracted from $1p-1s$ splitting of charmonium in ref.2($234MeV$).
\par
\vskip 10mm

In this work, the scaling behavior of SU(3) lattice theory in the
interval $\beta=5.85-7.60$
is studied by MCRG analysis.
A significant deviation from naive 2-loop scaling
is seen.
This is clearly observed in the data by measurements kept in the
confinement phase
up to $\beta=6.80$. Thus naive 2-loop  scaling does not hold
in this region.
It is shown that a large part of the deviation is accounted for by the
mean field
and/or the effective coupling constant schemes.  We show further that
the deviation can be absorbed by a next-order correction to the 2-loop
$\beta$-function, together with a shifted mean field coupling constant.
The coefficient of the next order term,$b'$, is consistently small.
Based on the scaling
behavior of this lattice $\beta$-function, the string tension remains constant
in the region $\beta=5.70-6.80$ where the lattice spacing changes by a factor
six.
The estimated $\Lambda_{\overline {MS}}^{n_f=0}$ in the continuum limit
is $ 0.46(0.03)\sqrt{\sigma}$. Though there is still a $20\%$ discrepancy
between different methods\cite{ref2,ref6,ref22}, this seems an encouraging
confirmation of the approach to the continuum limit.
 It is also noted that the suggested coupling constant $g_u$
diverges at some value of
the bare coupling constant( $\beta \sim 4.5$). An interesting possibility
is that
this divergence is related with
the transition point from strong to weak coupling region studied by
Bhanot and Creutz in a space of couplings of Wilson action and
that of adjoint representation.\cite{ref23}\par
Finally the significance of MCRG measurements in confinement phase is stressed.
Above the deconfining transition point, the matching procedure
inevitably suffers large errors.
Although most of the data
measured above the transition are consistent, those in confinement phase
have higher quality and reliability. Study on larger lattice is apparently
preferred. In order to cover the region up to $\beta$
=7.5, we need a 64$^4$ lattice.
\par
\vskip 10mm

{\bf Acknowledgement}\par
The present calculations have been carried on the parallel computer AP1000
at the Fujitsu Parallel Computing Research Facilities.
We would like to acknowledge Dr.M.Ishii and members of the facilities
for warm hospitality and valuable advice.
\par

\pagebreak
{\bf Figure Captions}
\par
\begin{description}
\item[Fig.1]  Wison loops used in lattice matching.
\item[Fig.2]  Coupling shift $\Delta\beta$ measured in this work(a) and
       that in previous works(b). Solid curve shows 2-loop
       asymptotic scaling with bare coupling constant.

\item[Fig.3]Autocorrelation time $\tau$ on $16^4$ and $32^4$ lattices.
\item[Fig.4]Coupling shift $\Delta\beta$ on $32^4$ lattice in
different blocking level.
\item[Fig.5]Coupling shift reproduced by different coupling constant schemes.
       Notations are described in the text.
\item[Fig.6]Scaling of string tension in different coupling constant
schemes.
\end{description}

\end{document}